\documentclass[a4paper,11pt]{article}
\usepackage{pos}

\title{Quantifying corrections to the hadron resonance gas with lattice QCD}

\author[a]{Rene Bellwied}
\author[b]{Szabolcs Borsányi}
\author[b,c,d,e]{Zoltán Fodor}
\author[b,f]{Jana N. Guenther}
\author[d,g]{Sándor D. Katz}
\author[b]{Paolo Parotto}
\author[d]{Attila Pásztor}
\author*[d]{Dávid Pesznyák}
\author[a]{Claudia Ratti}
\author[d,e]{Kálmán K. Szabó}

\affiliation[a]{Department of Physics, University of Houston,\\
Houston, TX, 77204, USA}
\affiliation[b]{Department of Physics, Wuppertal University,\\
Gaussstr. 20, D-42119, Wuppertal, Germany}
\affiliation[c]{Department of Physics, Pennsylvania State University,\\
State College, PA 16801, USA}
\affiliation[d]{Institute for Theoretical Physics, ELTE Eötvös Loránd University,\\
Pázmány P. sétány 1/A, H-1117, Budapest, Hungary}
\affiliation[e]{Jülich Supercomputing Centre, Forschungszentrum Jülich,\\
D-52425 Jülich, Germany}
\affiliation[f]{Aix Marseille Université, Université de Toulon\\
CNRS, CPT, Marseille, France}
\affiliation[g]{MTA-ELTE Theoretical Physics Research Group,\\
Pázmány P. sétány 1/A, H-1117 Budapest, Hungary}

\emailAdd{rbellwie@central.uh.edu}
\emailAdd{borsanyi@uni-wuppertal.de}
\emailAdd{fodor@bodri.elte.hu}
\emailAdd{jguenther@uni-wuppertal.de}
\emailAdd{katz@bodri.elte.hu}
\emailAdd{parotto@uni-wuppertal.de}
\emailAdd{apasztor@bodri.elte.hu}
\emailAdd{david.pesznyak@gmail.com}
\emailAdd{cratti@central.uh.edu}
\emailAdd{k.szabo@fz-juelich.de}

\abstract{The hadron resonance gas (HRG) model and its extensions are often used to describe the hadronic phase of strongly interacting matter. In our work we use lattice-QCD simulations with temporal extents of $N_\tau=8,10$ and $12$ to quantify corrections to the ideal HRG. Firstly, we determine a number of subleading fugacity expansion coefficients of the QCD free energy via a two-dimensional scan on the imaginary baryon number chemical potential ($\mu_B$) - strangeness chemical potential ($\mu_S$) plane. Using the aforementioned coefficients, we also extrapolate ratios of baryon number and strangeness fluctuations and correlations to finite chemical potentials via a truncated fugacity expansion. Our results extrapolated along the crossover line $T_\mathrm{c}(\mu_B)$ at strangeness neutrality are able to reproduce trends of experimental net-proton fluctuations measured by the STAR Collaboration.}

\FullConference{%
 The 38th International Symposium on Lattice Field Theory, LATTICE2021
  26th-30th July, 2021
  Zoom/Gather@Massachusetts Institute of Technology
}

\begin{document}
\maketitle

\section{Introduction}

    \noindent
    During the recent decades, the mapping of the QCD phase diagram has been in the focus of attention for both theoretical and experimental research. Many aspects of QCD thermodynamics are known at zero baryon density \cite{crossoverT1, crossoverT2, crossoverT3}, however, at finite $\mu_B$ the theory still proves to be perplexing. In particular, first principle lattice QCD simulations at non-vanishing $\mu_B$ are still hindered by the sign problem, although there are methods to circumvent it, such as reweighting \cite{reweighting1, reweighting2, reweighting3, reweighting4, reweighting5, reweighting6, reweighting7}, Taylor expansion at zero chemical potential \cite{taylor1, taylor2, taylor3, taylor4, taylor5}, and extrapolation from purely imaginary chemical potential \cite{imag1, imag2, imag3, imag4, imag5, imag6, imag7, imag8, imag9, imag10}. In this study the last was implemented, which also involves analytic continuation, hence it is essential to make use of some physical insight regarding the functional form of a given observable as a function of $\mu$.\bigskip
  
    \noindent
    The ideal hadron resonance gas (HRG) and its extensions are frequently used in describing the hadronic phase of strongly interacting matter \cite{HRG1, HRG2}. The ideal HRG model treats the interacting gas of hadrons as a free gas of hadrons and resonances, hence only the attractive resonant interactions are accounted for. The model performs well at describing thermodynamic observables at zero chemical potential, however, at finite $\mu$ a number of discrepancies emerge. In principle, the HRG can be systematically improved via the $S$-matrix formalism if the appropriate scattering matrix elements are known \cite{imag6, HRGPlus}. Furthermore, it is also straightforward to extend the HRG using a kind of mean field model that takes into account the short-range repulsive and non-resonant interactions. The importance of the hard-core short-range repulsive interactions has been pointed out \cite{imag7, meanField}, which leads to a significant negative contribution to the fugacity expansion coefficients with baryon number two.\bigskip
    
    \noindent
    The HRG model is also heavily used in the interpretation of experimental data. Apart from being a non-critical baseline, the so-called thermal fits provide means to estimate the chemical freeze-out temperature and chemical potential in heavy-ion collision experiments \cite{exp1, exp2}. Precise description of the hadronic phase -- including the effect of interactions -- is important to test different scenarios in heavy-ion physics, such as a single freeze-out temperature versus different freeze-out temperatures for light and strange hadrons \cite{sequential}.\bigskip
   
    \noindent
    In this paper we present our findings for subleading fugacity expansion coefficients from first principle lattice QCD simulations. We perform calculations in the two-dimensional plane of purely imaginary baryo- and strangeness chemical potentials. The fugacity expansion coefficients could be regarded as Fourier coefficients in the imaginary values of chemical potentials. This way we are able to separate the different contributions to the thermodynamics by the baryon number \textit{and} strangeness quantum numbers of the hadrons. We hope that our results will not only improve our understanding of the discrepancies between the HRG model and lattice results, but will give a new edge in constructing more realistic phenomenological models as well. Upon procuring the aforementioned coefficients we used a truncated fugacity expansion to extrapolate baryon number and strangeness fluctuation ratios of experimental interest to finite baryochemical potentials on the phenomenologically relevant strangeness neutral line. On the crossover line our results approximately reproduce the different trends seen in the net-proton fluctuation ratios from the STAR experiment. This conference contribution is based on Ref. \cite{cikk}.

\section{QCD in the Grand Canonical Ensemble}
\subsection{Generalised Susceptibilities}

    \noindent
    The generalised susceptibilities of the different quark flavours are defined as partial derivatives of the grand canonical potential (or pressure $p$) with respect to the different chemical potentials. For the case of the conserved charges baryon number $B$, electric charge $Q$ and strangeness $S$, we have:
    
    \begin{align}
    \label{eq:qSus}
        \chi_{ijk}^{BQS}=\frac{\partial^{i+j+k}(p/T^4)}{\partial\hat{\mu}_B^i\partial\hat{\mu}_Q^j\partial\hat{\mu}_S^k}\;,\\\nonumber
    \end{align}
    
    \noindent
    where $T$ denotes the temperature and e.g. $\hat{\mu}_B=\mu_B/T$ is the dimensionless chemical potential of the baryon number. The chemical potentials $\hat{\mu}_Q$ and $\hat{\mu}_S$ are defined analogously for electric charge and strangeness. In this proceedings we use $\mu_Q=0$ and only consider fluctuations and correlations of baryon number and strangeness.
    
\subsection{Fugacity Expansion of the QCD Free Energy}

    \noindent
    Apart from the Taylor expansion, one can use a Laurent expansion in the fugacity parameters $\exp(\hat{\mu}_B)$ and $\exp(\hat{\mu}_S)$ near 1. Based on charge conjugation symmetry the Laurent series can be expressed as an expansion in hyperbolic cosine functions of the chemical potentials, hence
    
    \begin{align}
    \label{eq:sector}
        \frac{p(T,\hat{\mu}_B,\hat{\mu}_S)}{T^4}=\sum\limits_{j,k}P_{jk}^{BS}(T)\cosh(j\hat{\mu}_B-k\hat{\mu}_S)\;.\\\nonumber
    \end{align}
    
    \noindent
    The $P_{jk}^{BS}(T)$ coefficients are the fugacity coefficients or sector coefficients referring to the fact that they only get contributions from Hilbert subspaces of fixed $B=j$ and $S=k$ quantum numbers -- for examples see Table \ref{tab:sectors}.\bigskip
    
    \begin{table}[ht]
        \centering
        \caption{Examples of different hadronic states contributing to the different sector coefficients $P_{jk}^{BS}$.}
        \begin{tabular}{c||c|c|c|c|c|c|c|c}
            $(B, S)$ & (0, 0) & (1, 0) & (0, 1) & (1, 1) & (1, 2) & (2, 0) & (2, 1) & (2, 2) \\\hline\hline
            hadrons & $\pi,\rho,\eta$ & $p,\Delta$ & $K$ & $\Lambda,\Sigma$ & $\Xi$ & $p$-$p$ & $K$-$K$ & $p$-$\Lambda$, $p$-$p$-$K$
        \end{tabular}
        \label{tab:sectors}
    \end{table}
    \bigskip
    
    \noindent
    The ideal HRG model receives its main contributions from the sectors $(B,S)=(0,0),(0,1),(1,0)$, $(1,1)$,$(1,2)$ and $(1,3)$. Further coefficients for $B\geq1$ are close to zero, which is -- as we will see -- a clear deviation from the lattice results. At purely imaginary chemical potentials $\mu=i\mu^{\mathcal{I}}$, and with $\cosh(i\mu^\mathcal{I})=\cos(\mu^\mathcal{I})$ Eq. \eqref{eq:sector} becomes a Fourier series. By differentiating Eq. \eqref{eq:sector} with respect to the $\mu_B$ and $\mu_S$ one can derive the functional forms of the generalised susceptibilities -- to be used later in the fitting procedure.\bigskip
    
    \noindent
    The coefficients $P_{01}^{BS}, P_{10}^{BS}, P_{11}^{BS}, P_{12}^{BS}$ and $P_{13}^{BS}$ are sizeable even in the ideal HRG model, however, they also get contributions from interactions such as $\pi$-$N$ scattering for the sector $P_{10}^{BS}$ or from $\Lambda$-$N$ interaction for $P_{21}^{BS}$. The $P_{1-1}^{BS}$ sector is zero in the ideal HRG, but it can get contributions from e.g. the $N$-$K^+$ scattering. The $B=2$ sectors get contributions from various processes; e.g. $P_{20}^{BS}$ from $N$-$N$, $P_{21}^{BS}$ from $N$-$\Lambda/\Sigma$, $P_{22}^{BS}$ from $N$-$\Xi$ or $\Lambda/\Sigma$-$\Lambda/\Sigma$ and $P_{23}^{BS}$ from $N$-$\Omega$ or $\Lambda/\Sigma$-$\Xi$ interactions. The $P_{02}^{BS}$ and $P_{03}^{BS}$ sectors get contributions from two- and three-kaon scatterings respectively. The $B_\mathrm{max}=3$ sectors include three-baryon scattering processes with various strangeness contents.
    
\section{Lattice Setup}
    
    \noindent
    In our lattice setup a staggered fermion action with four steps of stout smearing -- with smearing parameter $\rho=0.125$ -- and a tree-level Symanzik-improved gauge action was used \cite{taylor5}. For the scale setting the pion decay constant was fixed to $f_\pi=130.41$ MeV. The used lattice spacings are $N_\tau\times N_x^3=8\times24^3, 10\times32^3$ and $12\times36^3$, which correspond to the $LT\approx3$ aspect ratio. During continuum extrapolation we assume linear scaling in $1/N_\tau^2$. For each lattice spacing we ran simulations at four different temperatures: $T=145$ MeV, 150 MeV, 155 MeV and 160 MeV. At each temperature and lattice spacing a two-dimensional scan of the $(\hat{\mu}_B^\mathcal{I},\hat{\mu}_S^\mathcal{I})$ plane was performed at points $\hat{\mu}_B^\mathcal{I}=i\pi/8$ with $i=0,1,\dots,15$ and $\hat{\mu}_S^\mathcal{I}=j\pi/8$ with $j=0,1,\dots,8$, which sums up to a total of 144 simulation points. In each $\mu^{\mathcal{I}}\neq0$ point a Rational Hybrid Monte Carlo stream with several thousand trajectories was simulated, from which every fifth configuration was evaluated to determine the fluctuation observables. The statistical errors are handled with the jackknife method. During the systematic error estimation different continuum \textit{ansatzes} and different truncations of the fugacity expansion were taken into account.

\section{Numerical results}

    \noindent
    In order to estimate the sector coefficients $P_{jk}^{BS}$ a linear correlated fit was carried out. At vanishing chemical potential the $\chi_{20}^{BS},\chi_{11}^{BS},\chi_{02}^{BS},\chi_{40}^{BS},\chi_{31}^{BS},\chi_{22}^{BS},\chi_{13}^{BS}$ and $\chi_{04}^{BS}$ susceptibilities were considered, while for the remaining 143 simulated ensembles the $\mathrm{Im}\chi_{01}^{BS}$ and $\mathrm{Im}\chi_{10}^{BS}$ were used. The block diagonal covariance matrix was estimated with the jackknife method. The truncation of the fugacity expansion is not unequivocal, hence we introduced two different cuts in the maximal baryon number as $B_\mathrm{max}=2$ or 3. The used sectors in the correlated fits are presented in Table \ref{tab:fit}.\bigskip
    
    \begin{table}[ht!]
        \centering
        \caption{The used sectors in the $B_{\text{max}}=2$ and 3 fits. The $B_{\text{max}}=3$ column includes those sectors which are added to the $B_{\text{max}}=2$ case upon extension, hence we can detemine $12(+4)$ coefficients in total.}
        \begin{tabular}{l||llllllllllll|llll}
            & \multicolumn{12}{c|}{$B_{\text{max}}=2$} & \multicolumn{4}{c}{$B_{\text{max}}=3$} \\\hline\hline
            {$B,S$} & \footnotesize{1,0} & \footnotesize{0,1} & \footnotesize{1,-1} & \footnotesize{1,1} & \footnotesize{1,2} & \footnotesize{1,3} & \footnotesize{2,0} & \footnotesize{2,1} & \footnotesize{2,2} & \footnotesize{2,3} & \footnotesize{0,2} & \footnotesize{0,3} & \footnotesize{3,0} & \footnotesize{3,1} & \footnotesize{3,2} & \footnotesize{3,3}
        \end{tabular}
        \label{tab:fit}
    \end{table}
    \bigskip
    
    \begin{figure}[ht!]
        \centering
        \includegraphics[width = 0.7\textwidth]{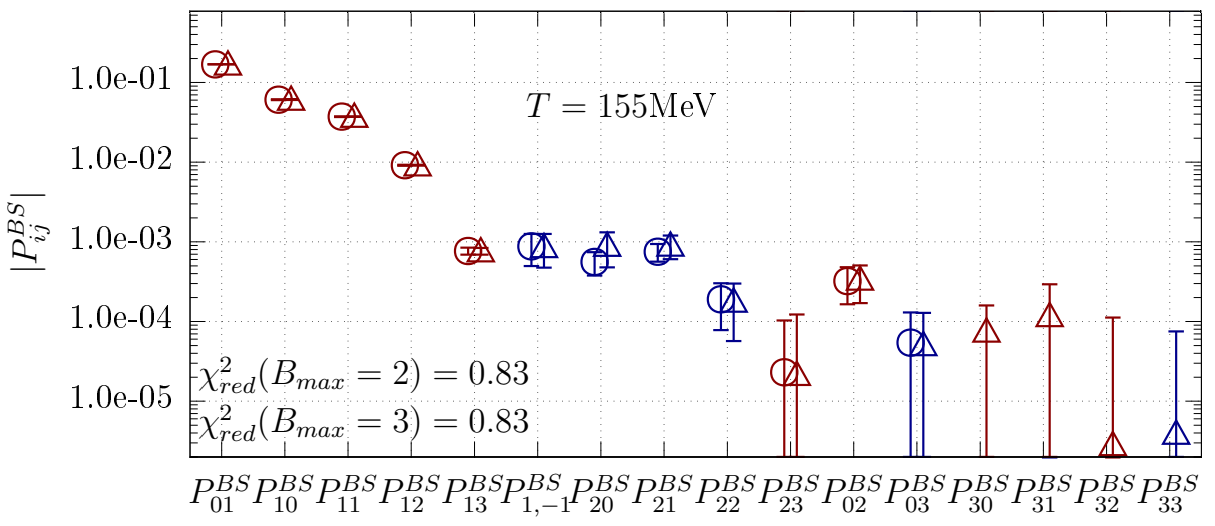}%
        \includegraphics[width = 0.15\textwidth]{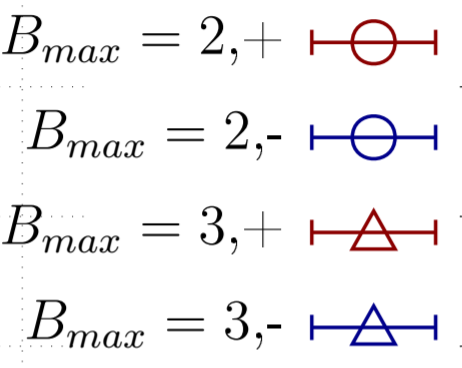}
        \caption{Results for the sector coefficients $P_{jk}^{BS}$ at $T=155$ MeV at our fines lattice. The leftmost five sectors are accounted for even in the ideal HRG model. For the next seven sectors we get stable solutions and are used later in phenomenological calculations. The $B_\mathrm{max}=3$ sectors are not stable enough as of yet. The vertical axis is set to log-scale, hence we introduced colours to indicate the sign of the coefficients; negative-blue, positive-red.}
        \label{fig:sectors}
    \end{figure}
    
    \noindent
    The results for the different sectors at our finest lattice $N_\tau\times N_x^3=12\times36^3$ at $T=155$ MeV are shown in Fig. \ref{fig:sectors}. It can be noted that the estimation of the $B=3$ sectors is not accurate enough, and their inclusion does not improve the quality of the fits, but the $B=2$ sectors remain consistent.\bigskip

    \noindent
    The continuum limit extrapolation of the sector coefficients proceeds through a combined two-dimensional fit with the following \textit{ansatz} as a function of temperature and temporal extent:
    
    \begin{align}
    \label{eq:continuum}
        f(T,N_\tau)=a_0+a_1T+a_2T^2+(b_0+b_1T+b_2T^2)\frac{1}{N_\tau^2}\;.\\\nonumber
    \end{align}
    
    \noindent
    In the systematic error estimation we have taken into account the effects whether we include the $B_\mathrm{max}=3$ set or not during the correlated fit and whether we include the coefficient $b_2$ in the continuum ansatz from Eq. \eqref{eq:continuum}. All the fits had good $\chi^2$ values. After the continuum limit fits are carried out we combine the four results with uniform weights for all the analysed sectors.\bigskip
    
\begin{figure}[ht!]
        \centering
        \includegraphics[width = 0.45\textwidth]{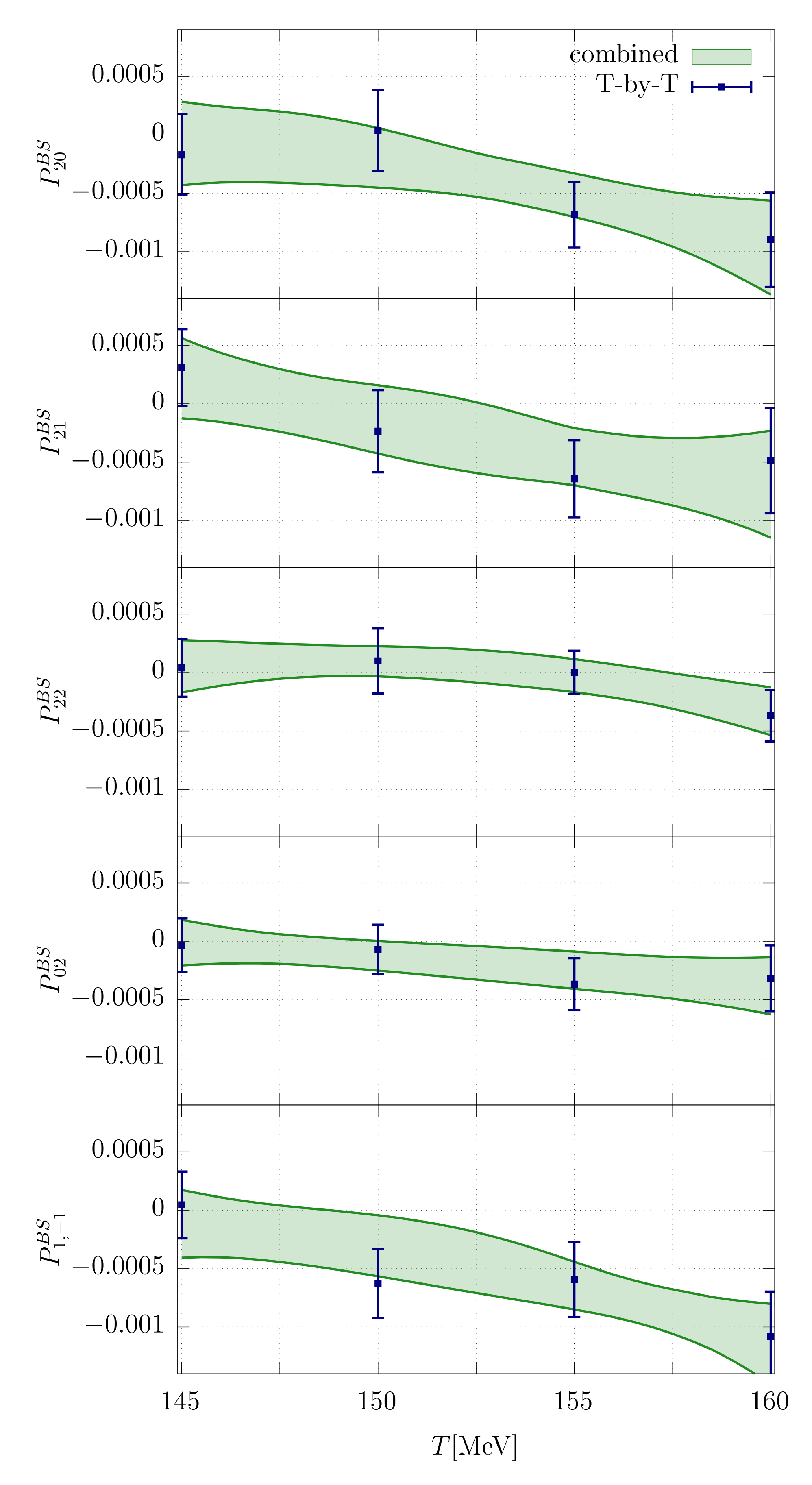}\hfill
        \includegraphics[width = 0.46\textwidth]{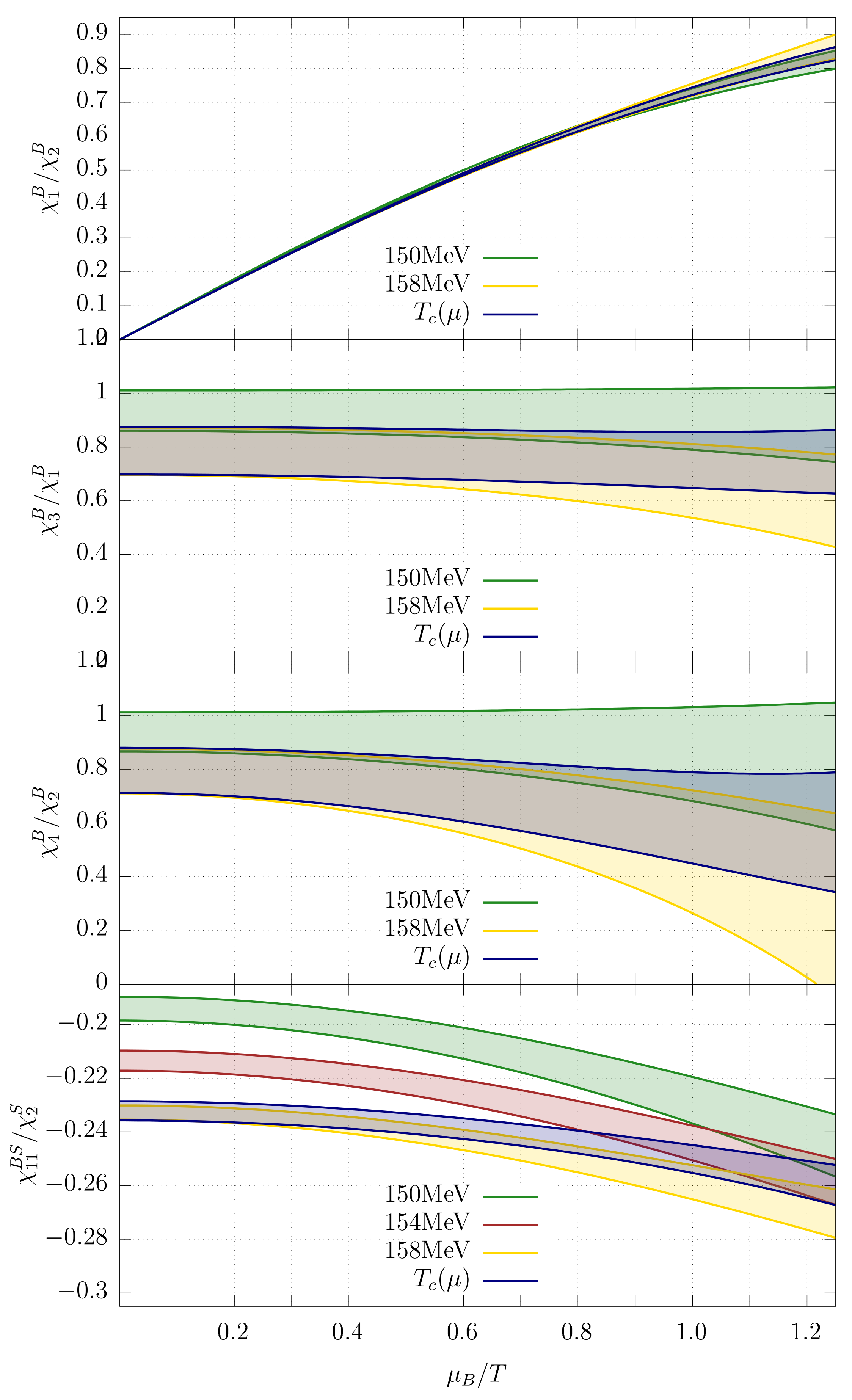}
        \caption{(Right) Our continuum limit estimates for the subleading sector coefficients from combined $T$-$1/N_\tau^2$ and $T$-by-$T$ continuum fits. Systematic errors coming from the choice of $B_\mathrm{max}$ and the inclusion of $b_2$ are taken into account. (Left) Continuum estimates of phenomenologically relevant generalised susceptibility ratios extrapolated to finite chemical potential at different temperatures and on the $T_\mathrm{c}(\mu_B)$ crossover line.}
        \label{fig:cont}
    \end{figure}

    \noindent
    The final results of our beyond-HRG sectors are shown in Fig. \ref{fig:cont}. According to our findings the sectors $P_{20}^{BS}$ and $P_{21}^{BS}$ are rather similar to each other. For higher temperatures they get more and more negative. This can be attributed to the short-range repulsive interactions, which start to take effect. The sector $P_{22}^{BS}$ gets smaller and for $P_{23}^{BS}$ we get a result roughly close to zero -- hence it is not included. As comparison, if one sums over the strangeness sectors in the ideal HRG as $\sum_kP_{2k}^{BS}$, the result will be $\mathcal{O}(10^{-5})$ -- orders of magnitude smaller than what we get from the lattice simulations. The sector $P_{02}^{BS}$ goes below zero within $1\sigma$ uncertainty at about 155 MeV. The results for $P_{1-1}^{BS}$ are rather large, which is in consistence with the phenomenological analysis in Ref. \cite{NK+}. Our results are also in approximate agreement with predictions from more phenomenological models. The sectors $P_{20}^{BS}$ and $P_{21}^{BS}$ are $1\sigma$ away from predictions of repulsive mean field models, while the $P_{22}^{BS}$ is within $2\sigma$ at all temperatures \cite{meanField}. Furthermore, the sector $P_{1-1}^{BS}$ is within $1\sigma$ agreement below 150 MeV and within $2\sigma$ at 160 MeV in $S$-matrix calculations \cite{NK+}. The leading order sector coefficients have already been published in earlier works \cite{imag6}, thus their analysis is not repeated here.\bigskip
    
    \noindent
    With the determined sector coefficients at hand different thermodynamic observables can be readily calculated. We consider the different baryon number and strangeness fluctuations and correlations at the experimentally relevant strangeness neutral line given by the constraint $\chi_{1}^{S}=0$. The extrapolation to real values of chemical potential is done via the expression in Eq. \eqref{eq:sector} at fixed $T$ and $N_\tau$ with truncation at $B_\mathrm{max}=2$, which is followed by the continuum limit estimation using the same \textit{ansatz} as for the sectors defined in Eq. \eqref{eq:continuum}. The same systematic error analysis was performed as for the fugacity expansion coefficients. The considered susceptibility ratios are $\chi_1^B/\chi_2^B$, $\chi_3^B/\chi_1^B$, $\chi_4^B/\chi_2^B$ and $\chi_{11}^{BS}/\chi_2^S$; our results are presented in Fig. \ref{fig:cont}. The parameterisation of the crossover line is $T_\mathrm{c}(\hat{\mu}_B)\approx T_\mathrm{c}^0(1-\kappa_2\hat{\mu}_B^2)$, where $T_\mathrm{c}^0=(158.0\pm0.6)$ MeV and $\kappa_2=0.0153\pm0.0018$ \cite{imag10}.\bigskip
    
\begin{figure}[ht!]
        \centering
        \includegraphics[width = 0.55\textwidth]{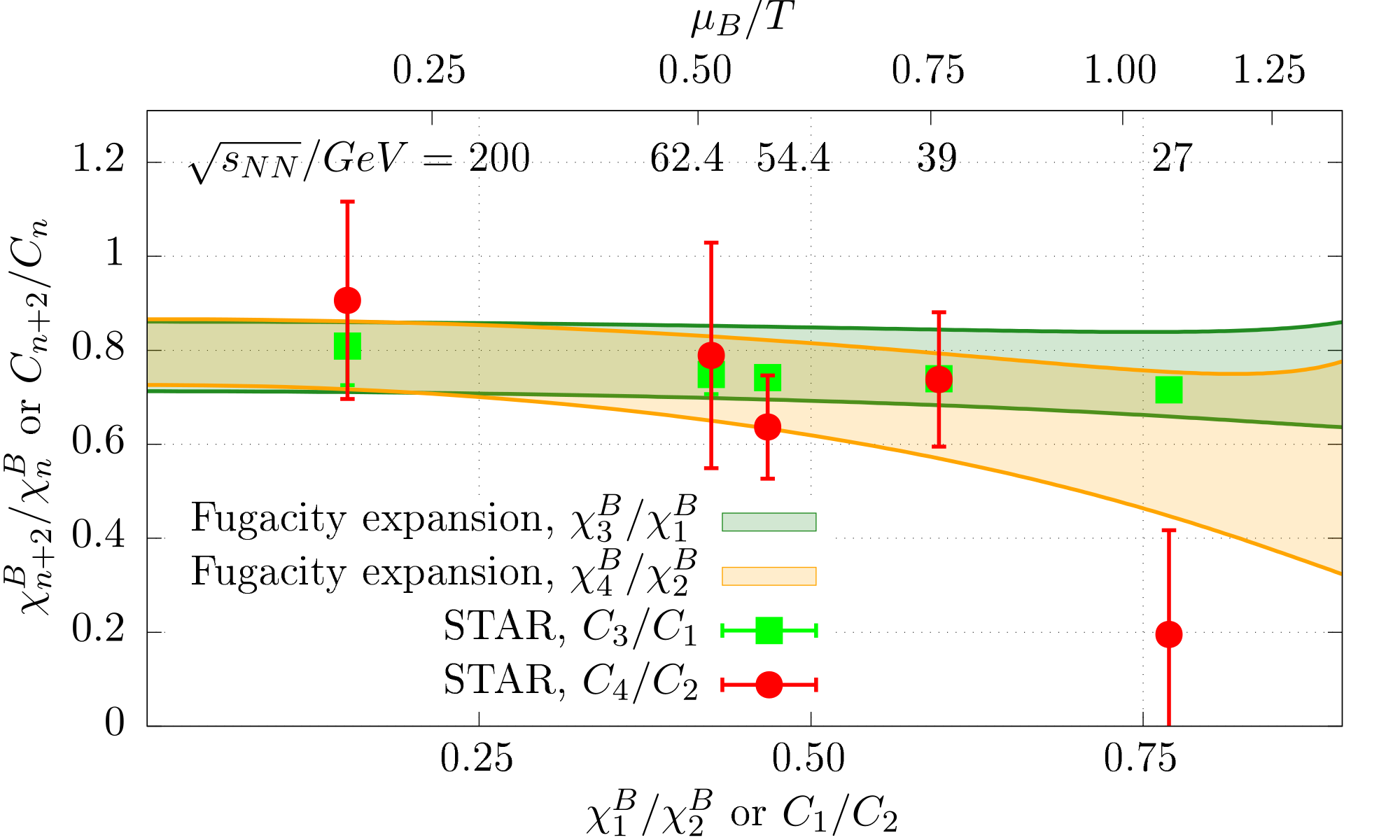}
        \caption{Comparison of our continuum estimates of fluctuation ratios with the STAR data.}
        \label{fig:STAR}
    \end{figure}

    \noindent
    The ratio $\chi_1^B/\chi_2^B$ shows strong dependence on the chemical potential, while depends weakly on the temperature, hence it can be used as a good proxy of $\mu_B$. The case for the ratios $\chi_4^B/\chi_2^B$, $\chi_3^B/\chi_1^B$ and $\chi_{11}^{BS}/\chi_2^S$ is quite the opposite, hence these ratios can be regarded as possible baryonic and strangeness thermometers. The latter ratio is of great phenomenological interest, since a good experimental proxy of it can be constructed from net-kaon and -lambda fluctuations as $\sigma_\Lambda^2/(\sigma_\Lambda^2+\sigma_K^2)$, for which it is shown that it is not strongly affected by experimental effects \cite{offDiagCorr}. Our results are consistent with recent lattice studies in which the ratios have been estimated using the Taylor method \cite{taylor5}. Furthermore, we also make comparisons with experimental net-proton skewness-to-mean $C_3/C_1$ and kurtosis-to-variance $C_4/C_2$ ratios from the STAR Collaboration \cite{STAR}. For this comparison one must assume that the chemical freeze-out and crossover lines are close to each other on the QCD phase diagram. The similarity of trends can be seen in Fig. \ref{fig:STAR}.\bigskip
    
\section{Summary and Outlook}

    \noindent
    In our present work we calculated the subleading sector coefficients of the QCD free energy from first principle lattice simulations. This way we could separate contributions coming from Hilbert subspaces with fixed baryon number and strangeness quantum numbers, hence the possibility is open to separate processes like $p$-$p$ or $K$-$K$ scatterings. We estimated the continuum limit results by considering lattices with temporal extents $N_\tau=8,10$ and 12 with aspect ratio $LT\approx3$. Our results quantify corrections to the ideal HRG directly from first principle calculations. After the estimation of the sector coefficients, we used a truncated fugacity expansion to calculate phenomenologically relevant fluctuation ratios. The comparison has many caveats \cite{caveat} but our results can reproduce the trends observed on the experimental STAR data of net-proton fluctuations.
    
\section*{Acknowledgements}

    \noindent
    This project was partly funded by the DFG Grant No. SFB/TR55 and also supported by the Hungarian National Research, Development and Innovation Office, NKFIH Grant No. KKP126769. The project leading to this publication has received funding from Excellence Initiative of Aix-Marseille University—A*MIDEX, a French ''Investissements d’Avenir'' programme, AMX-18-ACE-005. The project also received support from the Bundesministerium für Bildung und Forschung (BMBF) Grant No. 05P18PXFCA. Parts of this work were supported by the National Science Foundation under Grant No. PHY1654219 and by the U.S. Department of Energy (DOE), Office of Science, Office of Nuclear Physics, within the framework of the Beam Energy Scan Theory (BEST) Topical Collaboration. This research used resources of the Oak Ridge Leadership Computing Facility, which is a DOE Office of Science User Facility supported under Contract DE-AC05-00OR22725. A. P. is supported by the János Bolyai Research Scholarship of the Hungarian Academy of Sciences. A. P. and D. P. are supported -- respectively -- by the ÚNKP-21-5 and ÚNKP-21-2 New National Excellence Program of the Ministry of Innovation and Technology from the source of the National Research, Development and Innovation Fund. R. B. acknowledges support from the U.S. Department of Energy Grant No. DE-FG02-07ER41521. The authors gratefully acknowledge the Gauss Centre for Supercomputing (GCS) e.V. (\texttt{www.gauss-centre.eu}) for funding this project by providing computing time on the GCS Supercomputer JUWELS and JURECA/Booster at Jülich Supercomputing Centre (JSC). Parts of the computations were performed on the QPACE3 system, funded by the Deutsche Forschungsgemeinschaft (DFG). C. R. also acknowledges the support from the Center of Advanced Computing and Data Systems at the University of Houston.


\begin{thebibliography}{99}
\bibitem{crossoverT1}
Y.~Aoki, G.~Endrodi, Z.~Fodor, S.~D.~Katz and K.~K.~Szabo,
``The Order of the quantum chromodynamics transition predicted by the standard model of particle physics,''
Nature \textbf{443} (2006), 675-678

\bibitem{crossoverT2}
S.~Borsanyi \textit{et al.} [Wuppertal-Budapest],
``Is there still any $T_c$ mystery in lattice QCD? Results with physical masses in the continuum limit III,''
JHEP \textbf{09} (2010), 073

\bibitem{crossoverT3}
A.~Bazavov, T.~Bhattacharya, M.~Cheng, C.~DeTar, H.~T.~Ding, S.~Gottlieb, R.~Gupta, P.~Hegde, U.~M.~Heller and F.~Karsch, \textit{et al.}
``The chiral and deconfinement aspects of the QCD transition,''
Phys. Rev. D \textbf{85} (2012), 054503

\bibitem{reweighting1}
A.~Hasenfratz and D.~Toussaint,
``Canonical ensembles and nonzero density quantum chromodynamics,''
Nucl. Phys. B \textbf{371} (1992), 539-549

\bibitem{reweighting2}
Z.~Fodor and S.~D.~Katz,
``A New method to study lattice QCD at finite temperature and chemical potential,''
Phys. Lett. B \textbf{534} (2002), 87-92

\bibitem{reweighting3}
Z.~Fodor and S.~D.~Katz,
``Critical point of QCD at finite T and mu, lattice results for physical quark masses,''
JHEP \textbf{04} (2004), 050

\bibitem{reweighting4}
M.~Giordano, K.~Kapas, S.~D.~Katz, D.~Nogradi and A.~Pasztor,
``Effect of stout smearing on the phase diagram from multiparameter reweighting in lattice QCD,''
Phys. Rev. D \textbf{102} (2020) no.3, 034503

\bibitem{reweighting5}
M.~Giordano, K.~Kapas, S.~D.~Katz, D.~Nogradi and A.~Pasztor,
``New approach to lattice QCD at finite density; results for the critical end point on coarse lattices,''
JHEP \textbf{05} (2020), 088

\bibitem{reweighting6}
S.~Borsanyi, Z.~Fodor, M.~Giordano, S.~D.~Katz, D.~Nogradi, A.~Pasztor and C.~H.~Wong,
``Lattice simulations of the QCD chiral transition at real baryon density,''
\texttt{[arXiv:2108.09213 [hep-lat]]}.

\bibitem{reweighting7}
M.~Giordano, K.~Kapas, S.~D.~Katz, D.~Nogradi and A.~Pasztor,
``Radius of convergence in lattice QCD at finite $\mu_B$ with rooted staggered fermions,''
Phys. Rev. D \textbf{101} (2020) no.7, 074511

\bibitem{taylor1}
C.~R.~Allton, S.~Ejiri, S.~J.~Hands, O.~Kaczmarek, F.~Karsch, E.~Laermann, C.~Schmidt and L.~Scorzato,
``The QCD thermal phase transition in the presence of a small chemical potential,''
Phys. Rev. D \textbf{66} (2002), 074507

\bibitem{taylor2}
R.~V.~Gavai and S.~Gupta,
``QCD at finite chemical potential with six time slices,''
Phys. Rev. D \textbf{78} (2008), 114503

\bibitem{taylor3}
R.~Bellwied, S.~Borsanyi, Z.~Fodor, S.~D.~Katz, A.~Pasztor, C.~Ratti and K.~K.~Szabo,
``Fluctuations and correlations in high temperature QCD,''
Phys. Rev. D \textbf{92} (2015) no.11, 114505

\bibitem{taylor4}
M.~Giordano and A.~P\'asztor,
``Reliable estimation of the radius of convergence in finite density QCD,''
Phys. Rev. D \textbf{99} (2019) no.11, 114510

\bibitem{taylor5}
A.~Bazavov, D.~Bollweg, H.~T.~Ding, P.~Enns, J.~Goswami, P.~Hegde, O.~Kaczmarek, F.~Karsch, R.~Larsen and S.~Mukherjee, \textit{et al.}
``Skewness, kurtosis, and the fifth and sixth order cumulants of net baryon-number distributions from lattice QCD confront high-statistics STAR data,''
Phys. Rev. D \textbf{101} (2020) no.7, 074502

\bibitem{imag1}
P.~de Forcrand and O.~Philipsen,
``The QCD phase diagram for small densities from imaginary chemical potential,''
Nucl. Phys. B \textbf{642} (2002), 290-306

\bibitem{imag2}
M.~D'Elia and M.~P.~Lombardo,
``Finite density QCD via imaginary chemical potential,''
Phys. Rev. D \textbf{67} (2003), 014505

\bibitem{imag3}
P.~Cea, L.~Cosmai and A.~Papa,
``Critical line of 2+1 flavor QCD: Toward the continuum limit,''
Phys. Rev. D \textbf{93} (2016) no.1, 014507

\bibitem{imag4}
M.~D'Elia, G.~Gagliardi and F.~Sanfilippo,
``Higher order quark number fluctuations via imaginary chemical potentials in $N_f=2+1$ QCD,''
Phys. Rev. D \textbf{95} (2017) no.9, 094503

\bibitem{imag5}
P.~Alba, R.~Bellwied, S.~Borsanyi, Z.~Fodor, J.~G\"unther, S.~D.~Katz, V.~Mantovani Sarti, J.~Noronha-Hostler, P.~Parotto and A.~Pasztor, \textit{et al.}
``Constraining the hadronic spectrum through QCD thermodynamics on the lattice,''
Phys. Rev. D \textbf{96} (2017) no.3, 034517

\bibitem{imag6}
V.~Vovchenko, A.~Pasztor, Z.~Fodor, S.~D.~Katz and H.~Stoecker,
``Repulsive baryonic interactions and lattice QCD observables at imaginary chemical potential,''
Phys. Lett. B \textbf{775} (2017), 71-78

\bibitem{imag7}
C.~Bonati, M.~D'Elia, F.~Negro, F.~Sanfilippo and K.~Zambello,
``Curvature of the pseudocritical line in QCD: Taylor expansion matches analytic continuation,''
Phys. Rev. D \textbf{98} (2018) no.5, 054510

\bibitem{imag8}
S.~Borsanyi, Z.~Fodor, J.~N.~Guenther, S.~K.~Katz, K.~K.~Szabo, A.~Pasztor, I.~Portillo and C.~Ratti,
``Higher order fluctuations and correlations of conserved charges from lattice QCD,''
JHEP \textbf{10} (2018), 205

\bibitem{imag9}
S.~Borsanyi, Z.~Fodor, J.~N.~Guenther, R.~Kara, S.~D.~Katz, P.~Parotto, A.~Pasztor, C.~Ratti and K.~K.~Szabo,
``QCD Crossover at Finite Chemical Potential from Lattice Simulations,''
Phys. Rev. Lett. \textbf{125} (2020) no.5, 052001

\bibitem{imag10}
A.~P\'asztor, Z.~Sz\'ep and G.~Mark\'o,
``Apparent convergence of Pad\'e approximants for the crossover line in finite density QCD,''
Phys. Rev. D \textbf{103} (2021) no.3, 034511

\bibitem{HRG1}
R.~Dashen, S.~K.~Ma and H.~J.~Bernstein,
``$S$ Matrix formulation of statistical mechanics,''
Phys. Rev. \textbf{187} (1969), 345-370

\bibitem{HRG2}
R.~Venugopalan and M.~Prakash,
``Thermal properties of interacting hadrons,''
Nucl. Phys. A \textbf{546} (1992), 718-760

\bibitem{HRGPlus}
P.~M.~Lo, B.~Friman, K.~Redlich and C.~Sasaki,
``$S$-matrix analysis of the baryon electric charge correlation,''
Phys. Lett. B \textbf{778} (2018), 454-458

\bibitem{meanField}
P.~Huovinen and P.~Petreczky,
``Hadron resonance gas with repulsive interactions and fluctuations of conserved charges,''
Phys. Lett. B \textbf{777} (2018), 125-130

\bibitem{exp1}
A.~Andronic, P.~Braun-Munzinger, K.~Redlich and J.~Stachel,
``Decoding the phase structure of QCD via particle production at high energy,''
Nature \textbf{561} (2018) no.7723, 321-330

\bibitem{exp2}
V.~Vovchenko, V.~V.~Begun and M.~I.~Gorenstein,
``Hadron multiplicities and chemical freeze-out conditions in proton-proton and nucleus-nucleus collisions,''
Phys. Rev. C \textbf{93} (2016) no.6, 064906

\bibitem{sequential}
F.~A.~Flor, G.~Olinger and R.~Bellwied,
``Flavour and Energy Dependence of Chemical Freeze-out Temperatures in Relativistic Heavy Ion Collisions from RHIC-BES to LHC Energies,''
Phys. Lett. B \textbf{814} (2021), 136098

\bibitem{cikk}
R.~Bellwied, C.~Ratti, S.~Borsanyi, P.~Parotto, Z.~Fodor, J.~N.~Guenther, S.~D.~Katz, A.~Pasztor, D.~Pesznyak and K.~K.~Szabo,
``Corrections to the hadron resonance gas from lattice QCD and their effect on fluctuation-ratios at finite density,''
Phys. Rev. D \textbf{104} (2021) no.9, 094508

\bibitem{NK+}
C.~Fern\'andez-Ram\'\i{}rez, P.~M.~Lo and P.~Petreczky,
``Thermodynamics of the strange baryon system from a coupled-channels analysis and missing states,''
Phys. Rev. C \textbf{98} (2018) no.4, 044910

\bibitem{offDiagCorr}
R.~Bellwied, S.~Borsanyi, Z.~Fodor, J.~N.~Guenther, J.~Noronha-Hostler, P.~Parotto, A.~Pasztor, C.~Ratti and J.~M.~Stafford,
``Off-diagonal correlators of conserved charges from lattice QCD and how to relate them to experiment,''
Phys. Rev. D \textbf{101} (2020) no.3, 034506

\bibitem{STAR}
J.~Adam \textit{et al.} [STAR],
``Nonmonotonic Energy Dependence of Net-Proton Number Fluctuations,''
Phys. Rev. Lett. \textbf{126} (2021) no.9, 092301

\bibitem{caveat}
A.~Bzdak, S.~Esumi, V.~Koch, J.~Liao, M.~Stephanov and N.~Xu,
``Mapping the Phases of Quantum Chromodynamics with Beam Energy Scan,''
Phys. Rept. \textbf{853} (2020), 1-87

\end{thebibliography}
\end{document}